\documentclass[prd,preprint,showpacs]{revtex4}
\usepackage{amsmath}
\usepackage{latexsym}
\usepackage{amsfonts}
\usepackage{graphicx}
\usepackage{subfigure}
\begin{document}
\title{The dynamical behavior of the Extended Holographic Dark Energy with Hubble Horizon}
\author{Jie Liu}
\email{sxtyliujie@126.com}

\author{Yungui Gong}
\email{gongyg@cqupt.edu.cn}
\author{Ximing Chen}
\email{chenxm@cqupt.edu.cn}
\affiliation{ College of Mathematics and Physics, Chongqing University of Posts and Telecommunications, Chongqing 400065, China}
\begin{abstract}
The extended holographic dark energy model with the Hubble horizon as the infrared cutoff avoids the problem
of the circular reasoning of the holographic dark energy model. Unfortunately, it is hit with the no-go theorem.
In this paper, we consider the extended holographic dark energy model with a potential, $V(\phi)$,
for the Brans-Dicke scalar field. With the addition of a potential for the Brans-Dicke scalar field, the extended holographic dark energy
 model using the Hubble horizon as the infrared cutoff is a viable dark energy model, and the model has the dark energy dominated attractor solution.
\end{abstract}
\preprint{0912.0802}
\date{\today}
\maketitle

\section{Introduction}

Recent astronomical observations have provided strong evidence that we live in an accelerating universe \cite{acc1,acc2}. In most models, the accelerating expansion is driven by an exotic energy component with negative pressure called dark energy. The simplest dark energy model, which is consistent with current observations, is the cosmological constant model ($\Lambda$CDM). But the $\Lambda$CDM model is faced with the so-called cosmological constant problem: why is the energy of the vacuum so much smaller than what we estimate it should be? Therefore, lots of dynamical dark energy models were also proposed. For a review of dark energy models, see \cite{acc3}.

 One of the interesting dynamical dark energy models is the holographic dark energy (HDE) model which is based on the holographic principle.
 The holographic principle was thought as a fundamental principle of quantum gravity, and therefore
 may shed some lights on dark energy problem. The HDE model is derived from the relationship between the ultraviolet (UV) and the infrared (IR) cutoffs proposed by Cohen, Kaplan and Nelson in \cite{acc4}. Because of the limit set by the formation of a black hole (BH), the UV-IR relationship gives an upper bound on the zero point energy density $\rho_{h}=3L^{-2}/(8\pi G)$, which means that the maximum entropy of the system is of the order of $S^{3/4}_{\textrm{BH}}$. Here $L$ is the scale of IR cutoff
 and $S_{\textrm{BH}}$ is the entropy of black holes. However, the original HDE model with the Hubble scale as the IR cutoff failed to explain the accelerating expansion \cite{acc5}. Li solved the problem by discussing the possibilities of the particle and event horizons as the IR cutoff, and he found that only the event horizon identified as the IR cutoff leads to a viable dark energy model \cite{acc6}. The HDE model using the event horizon as the IR cutoff was soon found to be consistent with the observational data in \cite{acc7}. By considering the interaction between dark energy and matter in the HDE model with the event horizon as the IR cutoff, it was shown that the interacting HDE model realized the phantom crossing behavior \cite{acc8}. Other discussions on the HDE model can be found in \cite{acc9,acc10,acc11,acc12,acc13,acc14,acc15,acc16,acc17,acc18}.

The existence of the event horizon means that the Universe must experience accelerated expansion, so the HDE model with the event horizon as the IR cutoff faces the problem of circular reasoning. If the Hubble horizon can be used as the IR cutoff in the HDE model, then the HDE model is more successful and interesting. To achieve that, we need to go beyond Einstein's general relativity (GR). GR is very well tested only in the solar system, so it is reasonable to wonder whether GR is also valid in the ultralarge length scales. It was claimed that the current accelerating expansion of the Universe might well be the evidence of modifying GR. These models include the $1/R$ gravity \cite{acc19}, the $f(R)$ gravity \cite{acc20},the DGP model \cite{acc21}, and string inspired models \cite{acc22}. In addition to the vector and tensor fields describing the fundamental forces, there may exist scalar field. The simplest alternative to GR which includes a scalar field in addition to the tensor field is Brans-Dicke theory. Therefore, it is interesting to discuss the HDE model in the framework of Brans-Dicke theory. That was first done by Gong in \cite{acc23}, and the model is called the extended holographic dark energy (EHDE) model. The EHDE model was also discussed in \cite{acc24,acc25,acc26,acc27,acc28}.
However, it was found that the EHDE model with the Hubble horizon as the IR cutoff fails to give a viable dark energy model \cite{acc23,acc29}. In this paper, we study the EHDE model with the Hubble horizon as the IR cutoff, and show that the model is a viable dark energy model, if the Brans-Dicke scalar field has a potential, $V(\phi)$.
The paper is organized as follows: the EHDE model with a potential for the Brans-Dicke scalar field is discussed in Sec. II, and we conclude the paper in Sec. III.

\section{EHDE model with a potential}

The Lagrangian for Brans-Dicke theory with a scalar field potential in the Jordan frame is given by
\begin{equation}
\label{1}
  L_{\textrm{BD}}=\frac{\sqrt{-g}}{16\pi}[\phi
R-\omega g^{\mu\nu}\frac{\partial_{\mu}\phi \partial_{\nu}\phi}{\phi}-V(\phi)]-L_{m}(\psi, g_{\mu\nu}),
\end{equation}
where $\omega$ is the coupling constant; it recovers GR when $\omega\rightarrow \infty$.
The current observational constraint on $\omega$ is $\omega>10^4$ \cite{bdlimit1,bdlimit2}. By varying the action,
we get Einstein equation,
\begin{equation}
\label{2}
 R_{\mu\nu}-\frac{1}{2}g_{\mu\nu}R=\frac{\omega}{\phi^{2}}(\phi_{,\mu}\phi_{,\nu}-\frac{1}{2}
g^{\alpha\beta}\phi_{,\alpha}\phi_{,\beta}g_{\mu\nu})+\frac{1}{\phi}(\nabla_{\mu}\nabla_{\nu}\phi-g_{\mu\nu}\Box\phi)+\frac{8\pi}{\phi} T_{\mu\nu}-\frac{V(\phi)}{2\phi} g_{\mu\nu}.
\end{equation}
Based on the flat Friedmann-Robertson-Walker metric, we get the evolution equations from the Lagrangian (\ref{1}):
\begin{equation}
\label{3} H^2-\frac{\omega}{6}(\frac{\dot{\phi}}{\phi})^{2}+H\frac{\dot{\phi}}{\phi}-\frac{V}{6\phi}=\frac{8\pi}{3\phi}\rho,
\end{equation}
\begin{equation}
\label{4}
\ddot{\phi}+3H\dot{\phi}=\frac{8\pi}{3+2\omega}(\rho-3p)+\frac{2V-\phi V_{,\phi}}{3+2\omega},
\end{equation}
\begin{equation}
\label{5}
\dot{\rho}_h+3H(\rho_h+p_h)=0,
\end{equation}
\begin{equation}
\label{5a}
\dot{\rho}_m+3H\rho_m=0,
\end{equation}
where $V_{,\phi}(\phi)\equiv dV(\phi)/d\phi$ and the total energy density $\rho=\rho_{m}+\rho_{h}$.
In this model, the matter energy density $\rho_m$ and dark energy density
$\rho_h$ contribute to the total energy density,
and there is no interaction between dark energy and dark matter. Combining the above equations, we get
\begin{equation}
\frac{\ddot{a}}{a}=-\frac{8\pi}{3\phi} \frac{[3\omega p+(3+\omega)\rho]}{3+2\omega}-\frac{\omega}{3}(\frac{\dot{\phi}}{\phi})^2
+H\frac{\dot{\phi}}{\phi}-\frac{2V- \phi V_{,\phi}}{2\phi(3+2\omega)}+\frac{V}{6\phi},
\label{6}
\end{equation}
\begin{equation}
  \dot{H}=-\frac{8\pi}{\phi}\frac{[\omega(\rho+p)+2\rho]}{3+2\omega}-\frac{\omega}{2}(\frac{\dot{\phi}}{\phi})^{2}
  +2H \frac{\dot{\phi}}{\phi}-\frac{2V-\phi V_{,\phi}}{2\phi(3+2\omega)}.
\label{7}
\end{equation}
In Brans-Dicke theory, the scalar field $\phi$ takes the role of $1/G$, so the EHDE density with the Hubble horizon as the IR cutoff is
\begin{equation}
 \rho_{h}=\frac{3c^2\phi H^2}{8\pi},
\label{8}
\end{equation}
where $c$ is a constant of the order of unity. Using the dimensionless variables
\begin{equation}
x=\frac{\dot{\phi}}{H\phi}~,~~ y=\frac{1}{H}\sqrt{\frac{V}{6\phi}}~,~~\lambda=\frac{\phi V_{,\phi}}{V}~,~~
\Gamma=\frac{VV_{,\phi\phi}}{(V_{,\phi})^2},
\label{9}
\end{equation}
where $x\neq0~,~~y>0$, and $\lambda\neq 0$, Eq. (\ref{3}) becomes
\begin{equation}
\frac{8\pi}{3\phi H^2}\rho_{m}+c^2+\frac{\omega}{6} x^2-x+y^2=1.
\label{10}
\end{equation}
In this theory, the critical density is defined as $\rho_{c}=3\phi H^2/8 \pi$, so Eq. (\ref{10}) becomes
$\Omega_{m}+\Omega_{h}+\Omega_{\phi}=1$, where
\begin{equation}
\label{11}
\Omega_{m}=\frac{8\pi \rho_{m}}{3\phi H^2},\quad \Omega_{h}=c^2, \quad \Omega_{\phi}=\frac{\omega}{6}x^2-x+y^2 .
\end{equation}
From Eq. (\ref{11}), we see that the Brans-Dicke scalar field $\phi$ plays a role of dark energy,
so we assume that both the HDE and the scalar field drive our universe to accelerate, and
\begin{equation}
\Omega_{\textrm{de}}=\Omega_{h}+\Omega_{\phi}.
\label{12}
\end{equation}
Now
\begin{equation}
r=\frac{\Omega_{m}}{\Omega_{\textrm{de}}} =\frac{\Omega_{m}}{\Omega_{h}+\Omega_{\phi}}=\frac{1}{c^2+\frac{\omega}{6}x^2-x+y^2}-1.
\label{13}
\end{equation}
From Eq. (\ref{6}), we get the deceleration parameter
\begin{equation}
q=-\frac{\ddot{a}}{aH^2}=\frac{\omega(\omega+1)x^2-2\omega(1+c^2)x+3-6(\omega+\lambda)y^2}{2[3+2\omega(1-c^2)]}+\frac{1}{2}.
\label{14}
\end{equation}
With the definition (\ref{8}) of the EHDE and the energy conservation equation (\ref{5}), we get the equation of state parameter of the HDE:
\begin{equation}
w_{h}=\frac{3-(4\omega+3)x+\omega(\omega+1)x^2-6y^2(\lambda+\omega)}{3[3+2\omega(1-c^2)]},
\label{15}
\end{equation}
Combining Eqs. (\ref{5}), (\ref{11}) and (\ref{15}), we get the equation of state parameter of the Brans-Dicke scalar field,
\begin{equation}
\label{16}
w_{\phi}=\frac{w_{h}(1-c^2)}{\omega x^2/6-x+y^2}.
\end{equation}

Let $N=\ln a$, with the help of Eqs. (\ref{3}) and (\ref{7}), Eq. (\ref{4}) is rewritten as
\begin{equation}
\begin{split}
x'=&\frac{(\omega x-3)[(\omega+1)x-1]x}{2[3+2\omega(1-c^2)]}-\frac{3[(\omega+1)x-1][x+2(1-c^2)]}{2[3+2\omega(1-c^2)]} \\
&+\frac{3(3-\omega x)y^2}{3+2\omega(1-c^2)}
-\frac{3\lambda[x+2(1-c^2)]y^2}{3+2\omega(1-c^2)},
\end{split}\label{17}
\end{equation}
From the definition of the dimensionless variable and Eq.
(\ref{7}), we get
\begin{equation}
y'=\frac{1}{2}y(3+\lambda x)+\frac{(\omega
x-3)[(\omega+1)x-1]y}{2[3+2\omega(1-c^2)]}-\frac{3(\omega+\lambda)y^3}{3+2\omega(1-c^2)},
\label{18}
\end{equation}
From the definition of the dimensionless variable $\lambda$ , we get
\begin{equation}
\lambda'=x\lambda(1+\lambda\Gamma-\lambda)
\label{19},
\end{equation}
where $x'=dx/dN~,~y'=dy/dN$, and $\lambda'=d\lambda/dN$.

The system (\ref{17})-(\ref{19}) is not a three-dimensional autonomous system,
because the variable $\Gamma$ is unknown. However, if $ d \lambda /
dN=0$, considering  $x\neq0~,~~\lambda\neq0~$, we get
\begin{equation}
1+\lambda\Gamma-\lambda=0,
\label{20}
\end{equation}
Solving Eq. (\ref{20}), we get
\begin{equation}
V(\phi)=V_{0}\phi^{-n}~~(n\neq0),\quad \Gamma=\frac{n+1}{n},\quad \lambda=-n,
\label{21}
\end{equation}
where the power $n$ is a constant. Therefore if we consider the power-law potential $V(\phi)=V_{0}\phi^{-n}$ for the scalar field,
the system (\ref{17})-(\ref{19}) becomes a two-dimensional autonomous system:
\begin{equation}
\label{22}
x'=\frac{[(\omega+1)x-1][\omega x^2-6x-6(1-c^2)]}{2[3+2\omega(1-c^2)]}+
\frac{3[3+(n-\omega)x+2n(1-c^2)]y^2}{3+2\omega(1-c^2)},
\end{equation}
\begin{equation}
y'=\frac{1}{2}y(3-nx)+\frac{(\omega
x-3)[(\omega+1)x-1]y}{2[3+2\omega(1-c^2)]}-\frac{3(\omega-n)y^3}{3+2\omega(1-c^2)}.
\label{23}
\end{equation}
The fixed points of the autonomous system (\ref{22}) and (\ref{23}) are
\begin{equation}
\begin{split}
(~x_{c1}=\frac{1}{\omega+1},~y_{c1}=0~),\quad (~x_{c2}=\frac{3\pm\sqrt{3[3+2\omega(1-c^2)]}}{\omega},~y_{c2}=0~),\\
(~x_{c3}=\frac{3}{n},~y_{c3}=\frac{\sqrt{6(\omega+1)-2n}}{2n}~),\quad (~x_{c4},~y_{c4}=\sqrt{-\frac{\omega}{6}x_{c4}^{2}+x_{c4}+1-c^2}~),
\end{split}\label{24}
\end{equation}
where $x_{c4}=[4+2n(1-c^2)]/[2\omega+1-n]$. From the definition of the dimensionless variables (\ref{9}), we know the fixed points $(x_{c1},~y_{c1})$ and $(x_{c2},~y_{c2})$  are relevant to the system without a scalar field potential, and they are not the accelerated attractors. For the detailed discussion of them, see \cite{acc29}.

In this paper, we focus on the dynamical behavior of the fixed points $(x_{c3},~y_{c3})$, and $(x_{c4},~y_{c4})$.
In general, for an autonomous system
\begin{equation}
\begin{split}
x'=f(x,y),\quad y'=g(x,y),
\label{45}
\end{split}
\end{equation}
the nonsingular matrix at the fixed point ($x_{c},~y_{c}$) is,
\begin{displaymath}
\mathbf{M} =
\left( \begin{array}{ccc}
M_{11}=\frac{\partial f}{\partial x}(x_{c},y_{c}) & M_{12}=\frac{\partial f}{\partial y}(x_{c},y_{c}) \\
M_{21}=\frac{\partial g}{\partial x}(x_{c},y_{c}) & M_{22}=\frac{\partial g}{\partial y}(x_{c},y_{c})  \\
\end{array} \right).
\end{displaymath}
The eigenvalues of the matrix  are
\begin{equation}
\begin{split}
\frac{M_{11}+M_{22}\pm\sqrt{(M_{11}+M_{22})^2-4(M_{11}M_{22}-M_{12}M_{21})}}{2}.
\label{46}
\end{split}
\end{equation}
If the real parts of the eigenvalues of the matrix are negative, then the fixed point is a stable point.
So the stability conditions for the fixed point are
\begin{equation}
\begin{split}
M_{11}+M_{22}<0,~M_{11}M_{22}-M_{12}M_{21}>0.
\label{47}
\end{split}
\end{equation}

For the third fixed point $(x_{c3},~y_{c3})$, using Eq. (\ref{14}), we get the deceleration parameter $q=3/n+1/2$,
so it gives the accelerating expansion when $-6<n<0$. In order for the fixed point to be an accelerating attractor,
Eq. (\ref{47}) gives $n<-7(1+\sqrt{1+24(1+2\omega)/49})/4$,
$c^2<{\rm min}[(2n^2+7n-6\omega-3)/2n^2,~(3+2\omega)(1+1/n)/2(\omega+1)]$, and $\omega<4.5$
which is inconsistent with current observations.
Thus the stable fixed point $(x_{c3},~y_{c3})$ is not physical.

Now we analyze the fixed point $(x_{c4},~y_{c4})$. Combining Eqs. (\ref{11}) and (\ref{12}), we find
\begin{equation}
\Omega_{m}=0,\quad \Omega_{de}=\Omega_{h}+\Omega_{\phi}=1,
\label{25}
\end{equation}
and
\begin{equation}
r_{c4}=\frac{\Omega_{m}}{\Omega_{h}+\Omega_{\phi}}=0,
\label{26}
\end{equation}
so the fixed point $(x_{c4},~y_{c4})$ is relevant to a dark energy dominated universe.
To ensure the positivity of $\Omega_{\phi}$, we take $c^2<1$ .

With Eqs. (\ref{15})and (\ref{16}), we obtain
\begin{equation}
\label{27}
w_{\textrm{de}}=w_{\phi}=w_{h}=\frac{nx_{c4}}{3}-1=\frac{n[4+2n(1-c^2)]}{3(2\omega+1-n)}-1.
\end{equation}
Thus the HDE tracks the Brans-Dicke scalar field in the dark energy dominated era.

From Eq. (\ref{14}), the condition of acceleration becomes
\begin{equation}
q=\frac{(n+1)[2+n(1-c^2)]}{2\omega+1-n}-1<0.
\label{28}
\end{equation}

From Eq. (\ref{47}), the stability conditions for the fixed point  $(x_{c4}~,~y_{c4})$ are
\begin{gather}
\label{32}
2(M_{11}+M_{22})=3(nx_{c4}-3)-\frac{(3+2\omega)[x_{c4}+2(1-c^2)]}{[3+2\omega (1-c^2)]}-1<0,\\
\label{32a}
-2n(2\omega+1-n)x_{c4}+2n^2(1-c^2)+3+6\omega+n>0.
\end{gather}

\begin{figure}
$\begin{array}{cc}
\includegraphics[width=3in]{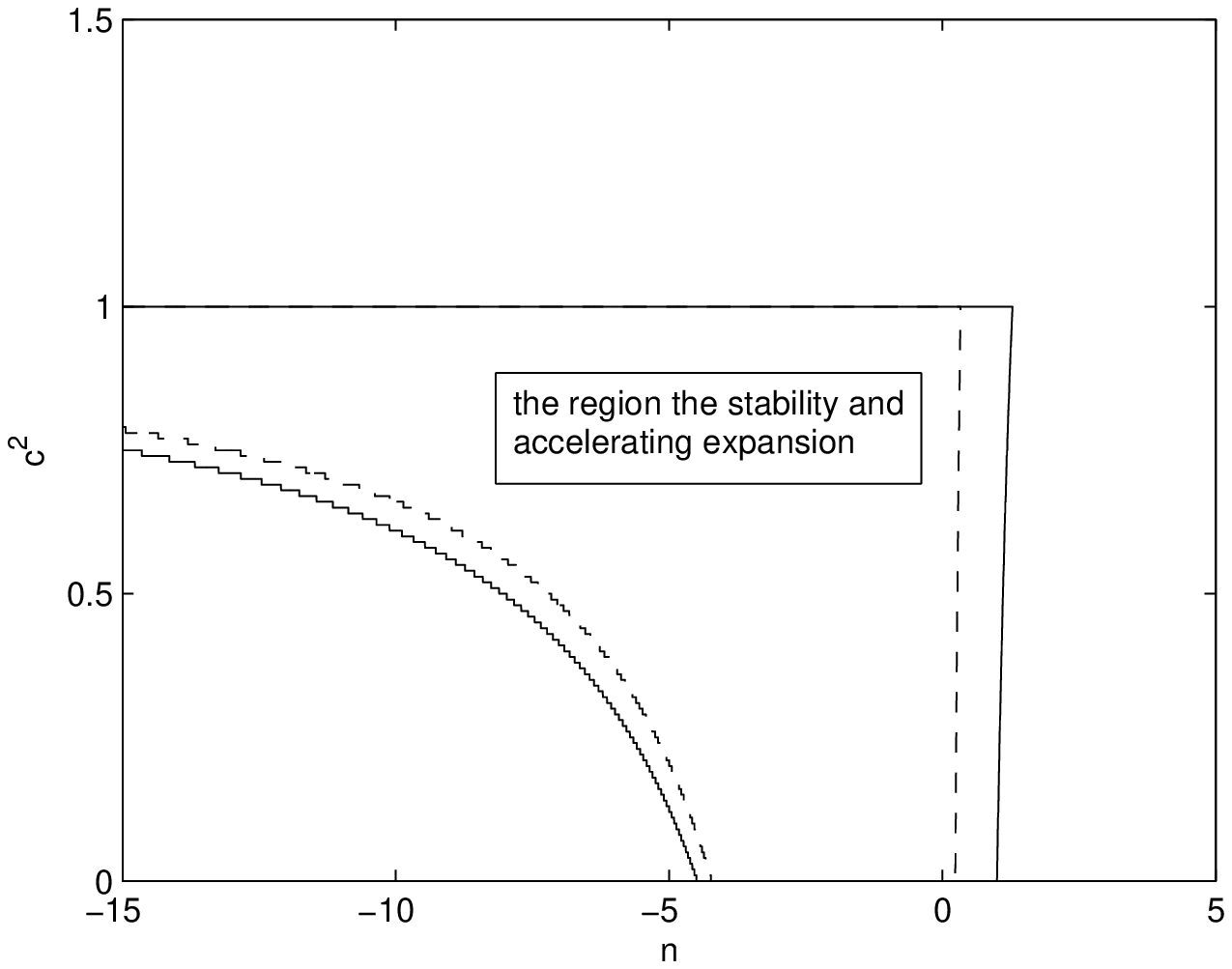} & \includegraphics[width=3in]{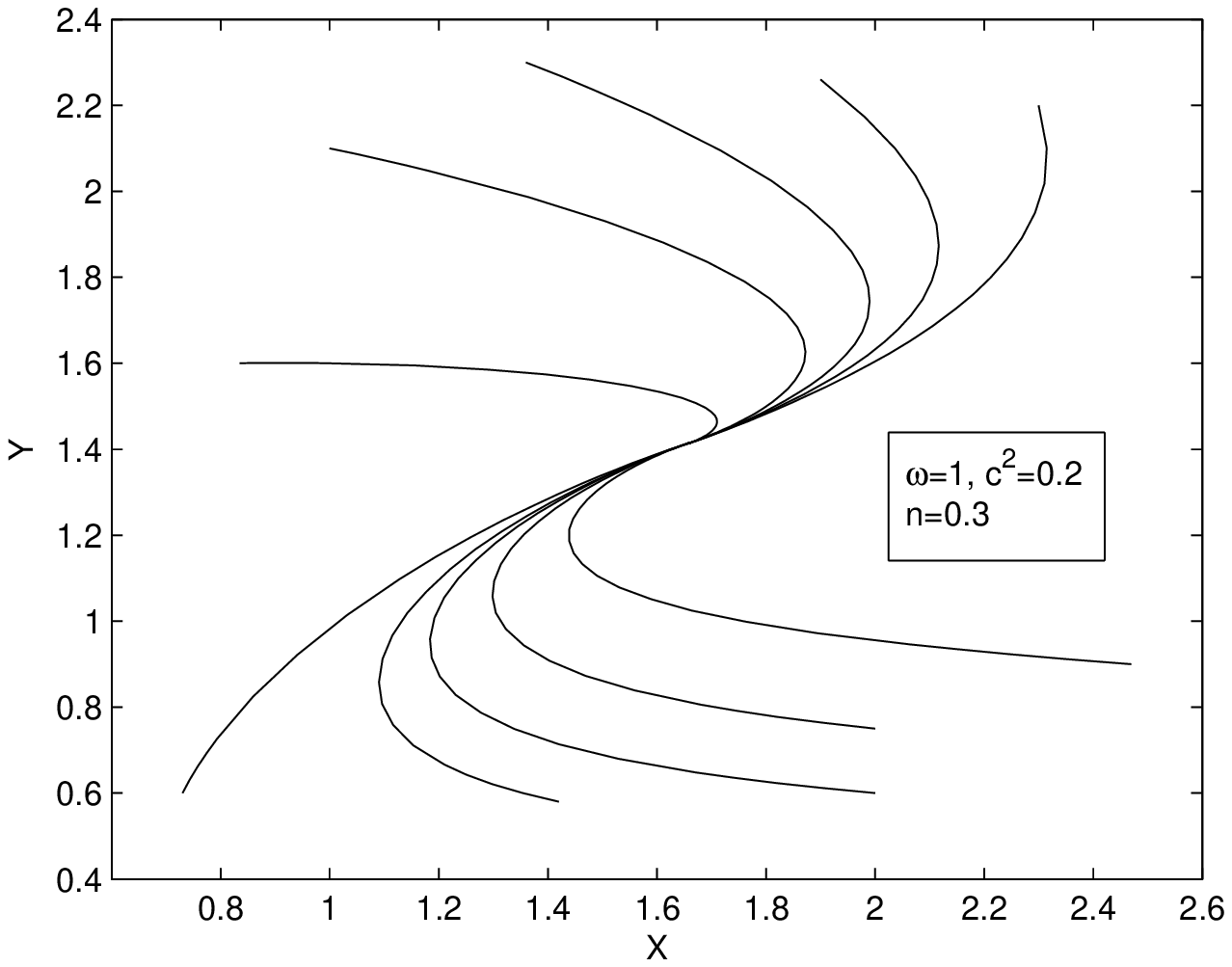}
\end{array}$
\caption{Left panel: The stability and accelerating expansion conditions for the fixed point $(x_{c4}~,~y_{c4})$ with $\omega=1$.
The solid line denotes the stability conditions and the dash-dot line denotes the acceleration conditions.
Right panel: Phase space trajectories for the fixed point $(x_{c4},~y_{c4})=(1.659~,~1.414)$, with $\omega=1$, $c^2=0.2$ and $n=0.3$.}
\label{fig1}
\end{figure}

\begin{figure}
$\begin{array}{cc}
\includegraphics[width=3in]{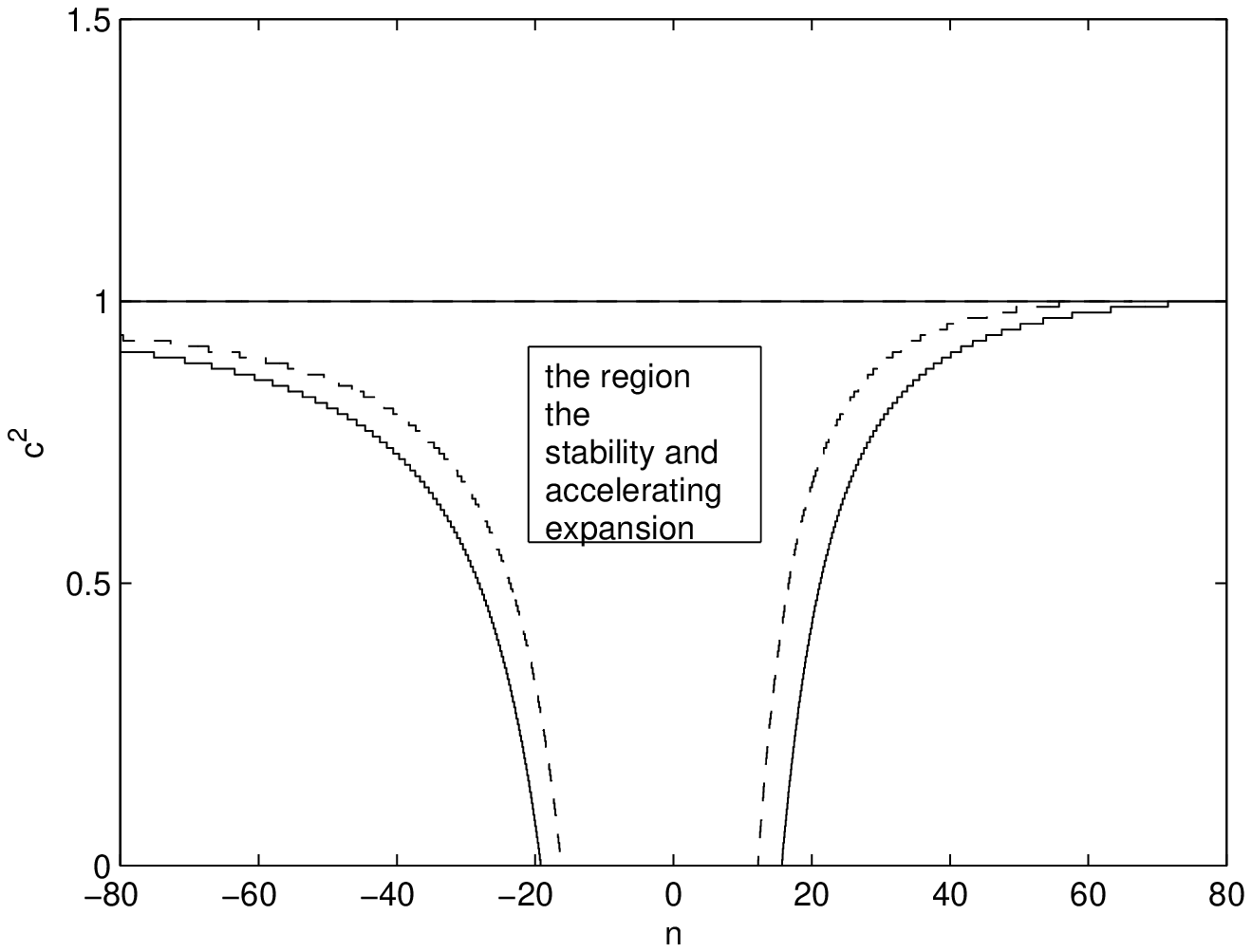} & \includegraphics[width=3in]{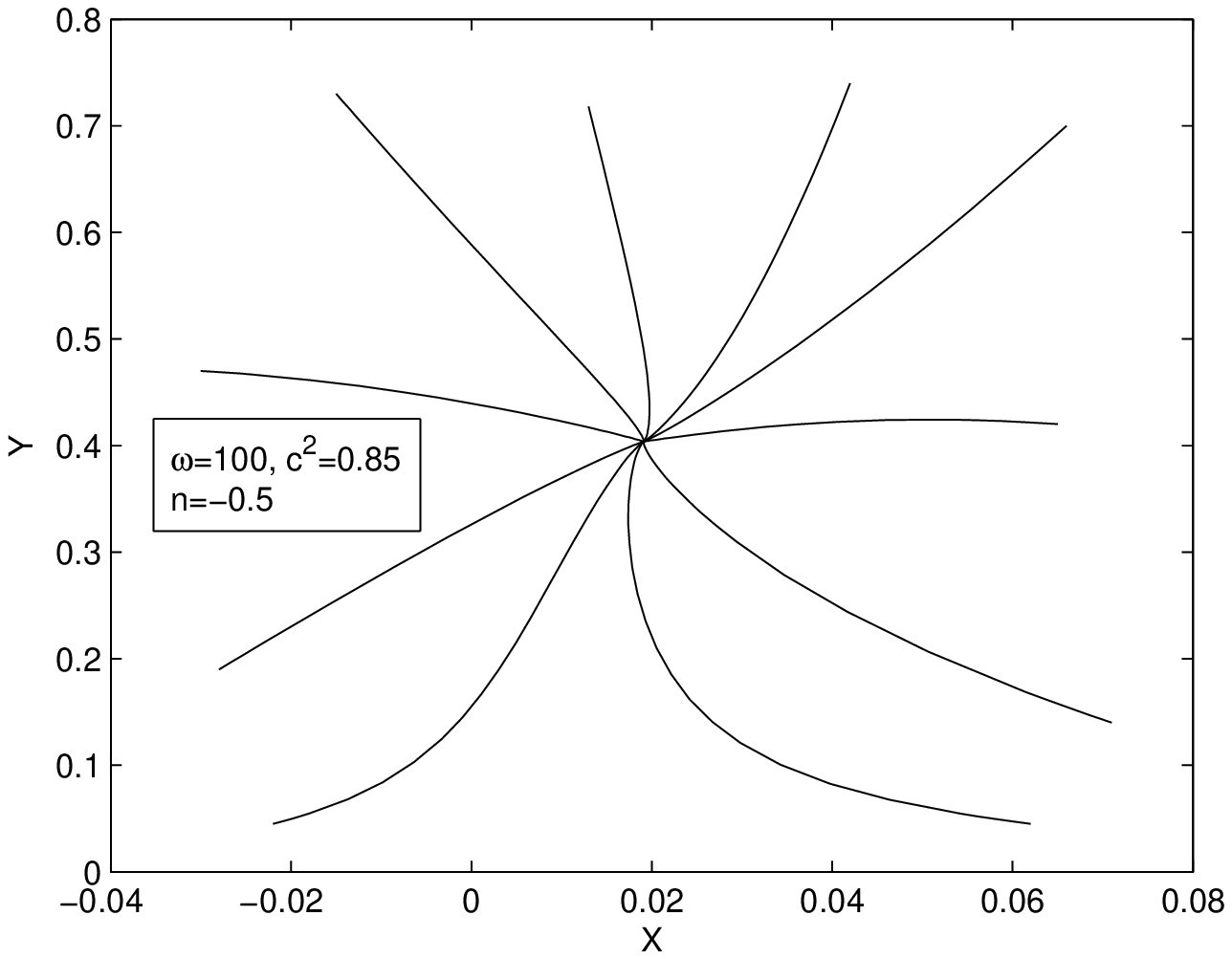}
\end{array}$
\caption{Left panel: The stability and accelerating expansion conditions for the fixed point $(x_{c4},~y_{c4})$ with $\omega=100$.
The solid line denotes the stability conditions and the dash-dot line denotes the acceleration conditions.
Right panel: Phase space trajectories for the fixed point $(x_{c4},~y_{c4})=(0.019,~0.404)$, with $\omega=100$, $c^2=0.85$ and $n=-0.5$.}
\label{fig2}
\end{figure}

\begin{figure}
$\begin{array}{cc}
\includegraphics[width=3in]{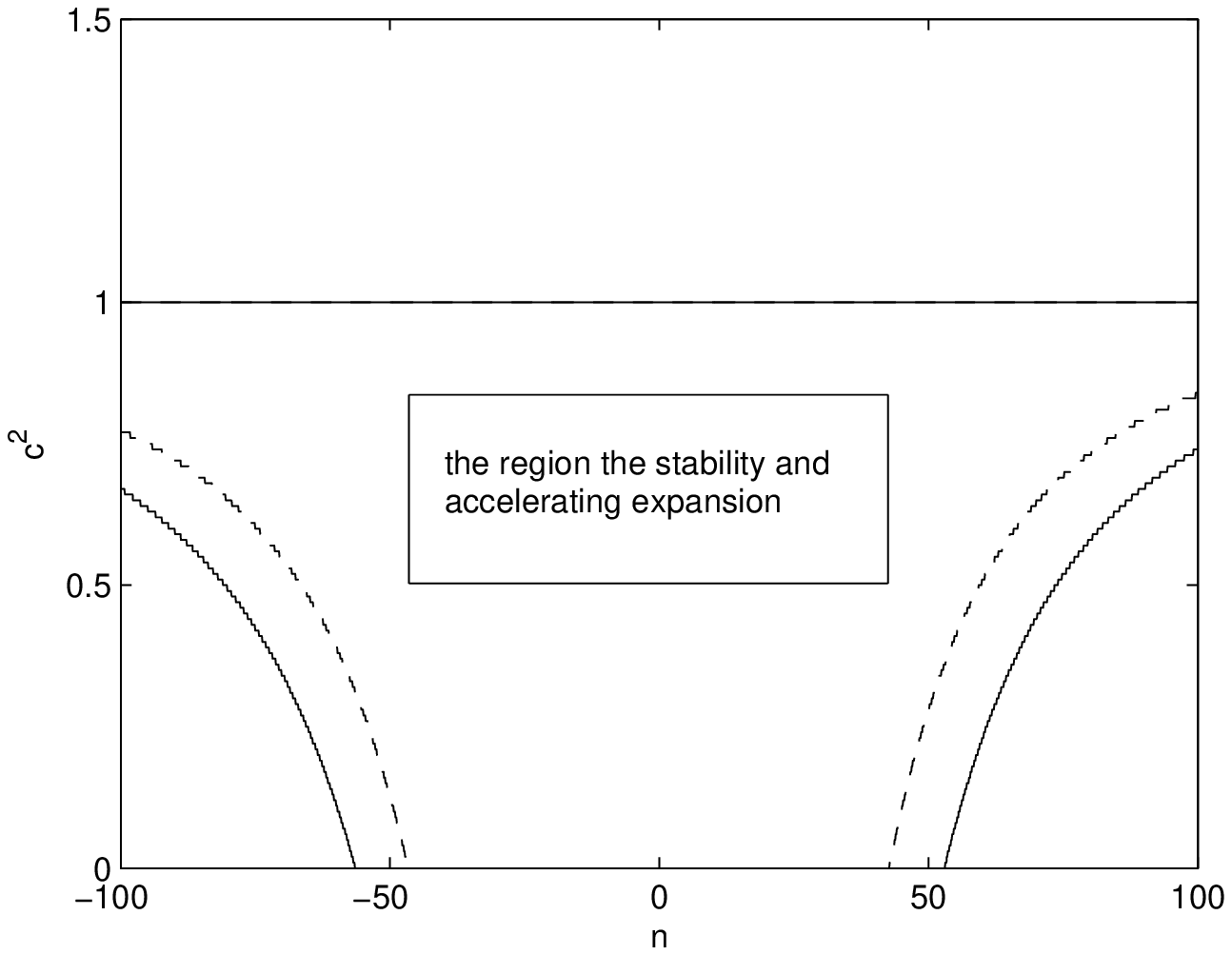} & \includegraphics[width=3in]{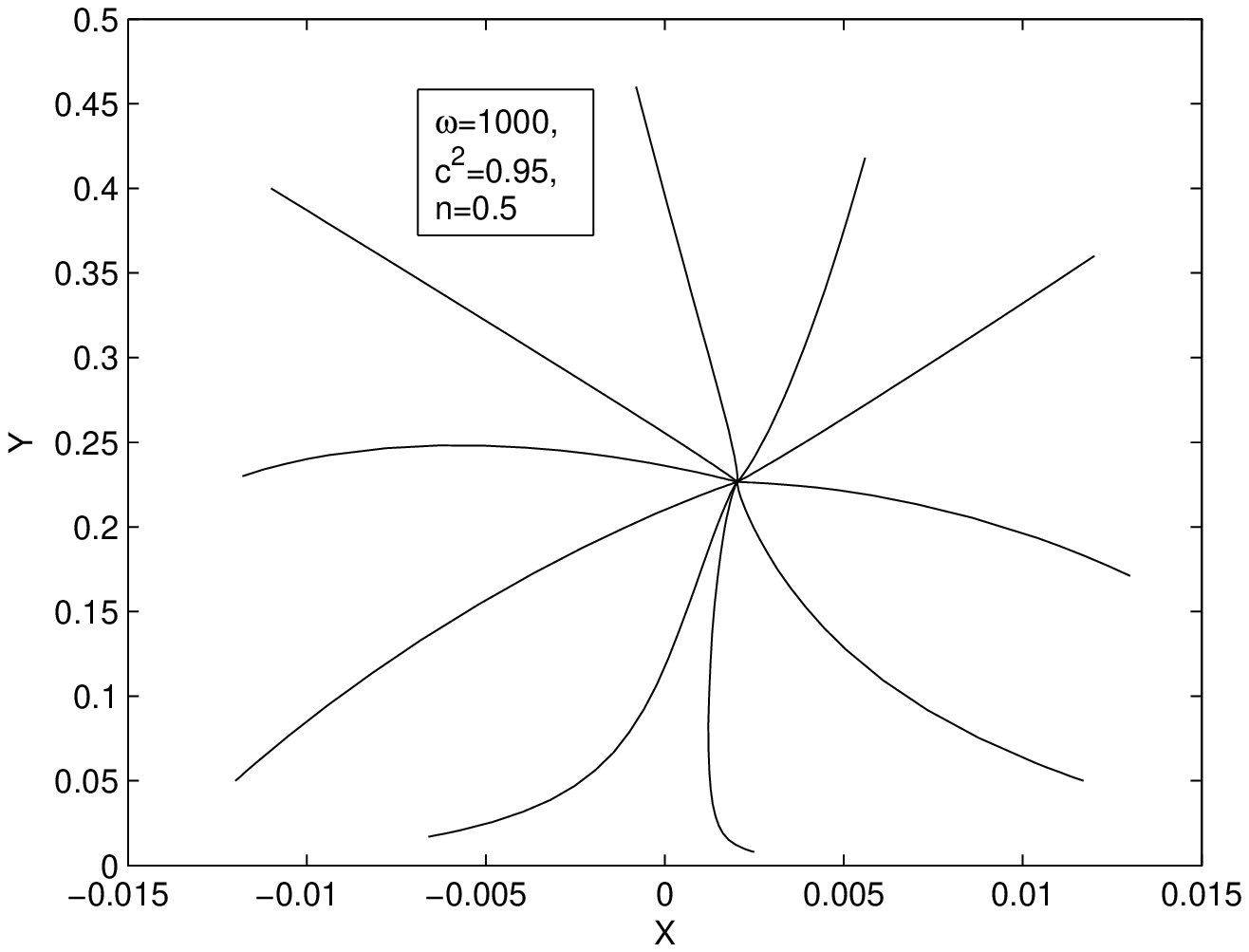}
\end{array}$
\caption{Left panel: The stability and accelerating expansion conditions for the fixed point $(x_{c4}~,~y_{c4})$ with $\omega=1000$.
The solid line denotes the stability conditions and the dash-dot line denotes the acceleration conditions.
Right panel: Phase space trajectories for the fixed point $(x_{c4},~y_{c4})=(0.002,~0.227)$, with $\omega=1000$, $c^2=0.95$ and $n=0.5$.}
\label{fig3}
\end{figure}

To find out the stability conditions for the fixed point  $(x_{c4},~y_{c4})$, we need to solve Eqs. (\ref{32}) and (\ref{32a})
numerically. In Eqs. (\ref{32}) and (\ref{32a}), there are three parameters $\omega$, $n$, and $c^2$.
From a theoretical point of view, the value of the Brans-Dicke parameter $\omega$ is expected to be order of unity. However, the current observational limit requires $\omega$ to be very large, so we choose some typical values, $\omega=1$, $\omega=100$, and $\omega=1000$, respectively, to find out the stable and accelerating regions in the parameter space $n$ and $c^2$.
The results are shown in Figs. \ref{fig1}, \ref{fig2} and \ref{fig3}.

The result of $\omega=1$ is plotted in Fig. \ref{fig1}. The left panel of Fig. \ref{fig1} shows the parameter
space for the fixed point $(x_{c4},~y_{c4})$  to be both stable and accelerated. To show the existence of the late time accelerating
attractors, as an example, we pick up the point $(c^2,~n)=(0.2,~0.3)$ that satisfies the stable and accelerated
conditions from the parameter space in the left
panel of Fig. \ref{fig1}, and the evolution of the phase space is shown in the right panel of Fig. \ref{fig1}.
With this choice of parameters, the corresponding fixed point $(x_{c4},~y_{c4})=(1.659,~1.414)$ is a late time accelerating attractor.
Using Eq. (\ref{26}), we get the dark energy equation of state parameter $w_{de}=-0.834$.
From Fig. \ref{fig1}, we see that whatever the initial conditions are, the universe can always evolve into the dark energy dominated state.

In Figs. \ref{fig2} and \ref{fig3}, we plot the stable and accelerated regions for $\omega=100$ and $\omega=1000$, respectively.
Again, we pick up a point from the stable and accelerated regions in the parameter space of $c$ and $n$ to
illustrate the property of the stable fixed point.
For $\omega=100$, we choose $(c^2,~n)=(0.85,~-0.5)$, and the evolution of the phase-space is shown in the right panel of Fig. \ref{fig2}.
With these parameters, the dark energy equation of state parameter $w_{de}=-1.003$.
The right panel of Fig. \ref{fig3} shows the evolution of the phase-space
with parameters $(c^2,~n)=(0.95,~0.5)$ and $\omega=1000$. With these parameters, the dark energy equation of state parameter $w_{de}=-0.9997$.
Since $\Omega_h=c^2=0.95$, the HDE is the dominate component of dark energy and the contribution from
the Brans-Dicke scalar field is negligible for driving the universe to accelerate.
From Figs. \ref{fig1}, \ref{fig2}, \ref{fig3}, we see that for any $\omega$, there exists some parameters of $c$ and $n$
so that the EHDE is a viable dark energy model, and the universe will evolve to the dark energy dominated state.
Furthermore, we find that
when the parameter $\omega$ becomes larger, the allowed region of the parameter $n$ becomes bigger, which means that
it is easier to get the dark energy dominated attractor.

\section{Discussion}

Note that when the parameter $\omega\rightarrow\infty$, Eq. (\ref{4}) becomes $\ddot{\phi}+3H\dot{\phi}\rightarrow 0$.
and the autonomous system (\ref{22}) and (\ref{23}) becomes
\begin{equation}
x'\simeq-\frac{3x(1-c^2+y^2)}{2(1-c^2)},
\label{29}
\end{equation}
\begin{equation}
y'\simeq\frac{1}{2}(3-nx)y-\frac{xy}{4(1-c^2)}-\frac{3y^3}{2(1-c^2)}.
\label{30}
\end{equation}

\begin{figure}
\centering
\includegraphics[width=3in]{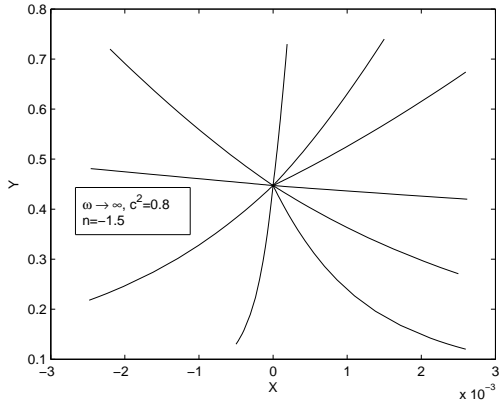}
\caption{Phase space trajectories for the fixed point$(x_{c4},~y_{c4})=(0,~0.4472)$, $w_{de}\rightarrow-1.0$ with $\omega\rightarrow\infty$, $c^2=0.8$ and $n=-1.5$}
 \label{fig4}
\end{figure}

The relevant fixed point of the autonomous system (\ref{29}) and (\ref{30}) is $(x_{c}=0,~y_{c}=\sqrt{1-c^2})$, and $c^2\neq1$.
The fixed point is always stable when $0<c^2<1$. Besides, from Eqs. (\ref{27}) and (\ref{28}), we get the equation of
state parameter of dark energy $w_{de}\simeq-1$ and the deceleration parameter $q\simeq-1$, so the system is accelerated.
We plot the phase space trajectories for the fixed point $(x_{c}=0,~y_{c}=\sqrt{1-c^2})$
with parameters $(c^2,~n)=(0.8,~-1.5)$ in Fig. \ref{fig4}. From Fig. \ref{fig4}, we see that the fixed point is a stable fixed point.

The above result is easily understood as follows. When $\omega\rightarrow\infty$, the Brans-Dicke scalar field becomes a constant,
and the potential becomes an effective cosmology constant.
Thus the Brans-Dicke theory with a potential for the Brans-Dicke scalar field reduces to the model
with a cosmological constant $\Lambda$ when $\omega\rightarrow\infty$, and the HDE becomes a
cosmological constant effectively.

The HDE model with the event horizon as IR cutoff has the problem of circular reasoning.
The EHDE model without a potential for the Brans-Dicke scalar field faces the same problem. In this paper,
we considered the EHDE model with a power-law potential and find that the model is a viable
dark energy model when the Hubble horizon is chosen as the IR cutoff. Therefore, the EHDE model
with a power-law potential avoids the problem of circular reasoning. With the addition of a
potential for the Brans-Dicke scalar field, the EHDE model with the Hubble horizon as the IR cutoff
has the dark energy dominated attractor solution. The Brans-Dicke scalar field partly serves as an
effective dark energy.

\begin{acknowledgments}
The work is supported by the National
Natural Science Foundation of China key project under Grant
No. 10935013, the Ministry of Science and
Technology of China national basic science program (973 Project)
under Grant No. 2010CB833004, and the Natural
Science Foundation Project of CQ CSTC under Grant No. 2009BA4050.
\end{acknowledgments}


\begin{thebibliography}{20}
\bibitem{acc1} A.G. Riess {\it et al.}, Astron. J. {\bf 116}, 1009 (1998).
\bibitem{acc2} S. Perlmutter {\it et al.}, Astrophy. J. {\bf 517}, 565 (1999).
\bibitem{acc3} V. Sahni and A.A. Starobinsky, Int. J. Mod. Phys. D {\bf 9}, 373 (2000);
T. Padmanabhan, Phys. Rep. {\bf 380}, 235 (2003); P.J.E. Peebles and B. Ratra, Rev. Mod. Phys. {\bf 75}, 559 (2003);
E.J. Copeland, M. Sami and S. Tsujikawa, Int. J. Mod. Phys. D {\bf 15}, 1753 (2006).
\bibitem{acc4} A. Cohen, D. Kaplan and A. Nelson, Phys. Rev. Lett. {\bf 82}, 4971 (1999).
\bibitem{acc5} S.D.H. Hsu, Phys. Lett. B {\bf 594}, 13 (2004).
\bibitem{acc6} M. Li, Phys. Lett. B {\bf 603}, 1 (2004).
\bibitem{acc7} Q.G. Huang and Y.G. Gong, J.Cosmol.Astropart.Phys. 08 (2004) 006.
\bibitem{acc8} B. Wang, Y.G. Gong and E. Abdalla, Phys. Lett. B {\bf 624}, 141 (2005).
\bibitem{acc9}Y.G. Gong, B. Wang and Y.Z. Zhang, Phys. Rev. D {\bf 72}, 043510 (2005);
H.-C. Kao, W.-L. Lee and F.-L. Lin, Phys Rev. D {\bf 71}, 123518 (2005);
Y.G. Gong and Y.Z. Zhang, Classical Quantum Grav. {\bf 22}, 4895 (2005).
\bibitem{acc10} Q. Wu, Y.G. Gong, A. Wang and J.S. Alcaniz, Phys. Lett. B {\bf 659}, 34 (2008).
\bibitem{acc11} Q.G. Huang and M. Li, J.Cosmol.Astropart.Phys. 08 (2004) 013; B. Chen, M. Li and Y. Wang, Nucl. Phys. B {\bf 774}, 256 (2007).
\bibitem{acc12} B. Wang, C.Y. Lin and E. Abdalla, Phys. Lett. B {\bf 637}, 357 (2006).
\bibitem{acc13} D. Pav\'{o}n and W. Zimdahl, Phys. Lett. B {\bf 628}, 206 (2005); M.R. Setare, Phys. Lett. B {\bf 642}, 1 (2006).
\bibitem{acc14} M. Ito, Europhys. Lett. {\bf 71}, 712 (2005); S. Nojiri and S.D. Odintsov, Gen. Rel. Grav. {\bf 38}, 1285 (2006).
\bibitem{acc15} B. Guberina, R. Horvat and H. Nikolic, Phys. Rev. D {\bf 72}, 125011 (2005); Phys. Lett. B {\bf 636}, 80 (2006).
\bibitem{acc16} B. Hu and Y. Ling, Phys. Rev. D {\bf 73}, 123510 (2006); H. Li, Z.K. Guo and Y.Z. Zhang, Int. J. Mod. Phys. D {\bf 15}, 869 (2006).
\bibitem{acc17} H.M. Sadjadi, J.Cosmol.Astropart.Phys. 02 (2007) 026; Z.K. Guo, N. Ohta and S. Tsujikawa, Phys. Rev. D {\bf 76}, 023508 (2007).
\bibitem{acc18} X. Zhang and F.Q. Wu, Phys. Rev. D {\bf 72}, 043524 (2005); {\bf 76}, 023502 (2007).
\bibitem{acc19}S.M. Carroll, V. Duvvuri, M. Trodden and M.S. Turner, Phys. Rev. D {\bf 70}, 043528 (2004); T. Chiba,
Phys. Lett. B {\bf 575}, 1 (2003); S. Nojiri and S.D. Odintsov, Phys. Rev. D {\bf 68}, 123512 (2003); C.G.
Shao, R.G. Cai, B. Wang and R.K. Su, Phys. Lett. B {\bf 633}, 164 (2006).
\bibitem{acc20}S. Capozziello, Int. J. Mod. Phys. D {\bf 11}, 483 (2002); S. Nojiri and S.D. Odintsov, Int. J. Geom.
Methods Mod. Phys. {\bf 4}, 115 (2007); W. Hu and I. Sawicki, Phys. Rev. D {\bf 76}, 064004 (2007).
\bibitem{acc21}G.R. Dvali, G. Gabadadze and M. Porrati, Phys. Lett. B {\bf 485}, 208 (2000); C. Deffayet, G.R. Dvali and
G. Gabadadze, Phys. Rev. D {\bf 65}, 044023 (2002); Y.G. Gong and C.K. Duan, Classical Quantum Grav.
{\bf 21}, 3655 (2004);Mon. Not. R. Astron. Soc. {\bf 352}, 847 (2004); Y.G.
Gong, Phys. Rev. D {\bf 78}, 123010 (2008).
\bibitem{acc22}P. Bin\'{e}truy, C. Deffayet and D. Langlois, Nucl. Phys. B {\bf 565}, 269 (2000); R.G. Cai, Y.G. Gong and B.
Wang, J.Cosmol.Astropart.Phys. 03 (2006) 006; Y.G. Gong and A. Wang, Class. Quantum Grav. {\bf 23}, 3419 (2006);
Y.G. Gong, A. Wang and Q. Wu, Phys. Lett. B {\bf 663}, 147 (2008).
\bibitem{acc23} Y.G. Gong, Phys. Rev. D {\bf 70}, 064029 (2004).
\bibitem{acc24} H. Kim, H.W. Lee and Y.S. Myung, Phys. Lett. B {\bf 632}, 605 (2006).
\bibitem{acc25} M.R. Setare, Phys. Lett. B {\bf 644}, 99 (2007).
\bibitem{acc26} N. Banerjee and D. pav\'{o}n, Phys. Lett. B {\bf 647}, 477 (2007).
\bibitem{acc27}B. Nayak, L.P. Singh, Mod. Phys. Lett. A {\bf 24}, 1785 (2009).
\bibitem{acc28} L. Xu and J. Lu, Eur. Phys. J. C {\bf 60}, 135 (2009).
\bibitem{acc29} Y.G. Gong and J. Liu, J.Cosmol.Astropart.Phys.09 (2008) 010.
\bibitem{bdlimit1} C.M. Will,  Living Rev. Rel. {\bf 9}, 3 (2006).
\bibitem{bdlimit2} B. Bertotti, L. Iess and P. Tortora,  Nature {\bf 425}, 374 (2003).

\end{thebibliography}
\end{document}